\documentclass[prd,eqsecnum,twocolumn,amsfonts,amssymb]{revtex4}
\usepackage{graphicx}
\graphicspath{{./figures/}}

\usepackage{bm}

\newcommand{\beq}{\begin{equation}}
\newcommand{\eeq}{\end{equation}}
\newcommand{\beqs}{\begin{eqnarray}}
\newcommand{\eeqs}{\end{eqnarray}}

\newcommand{\gsim}{\mathrel{\raisebox{-
.6ex}{$\stackrel{\textstyle>}{\sim}$}}}
\newcommand{\dslash}{\partial\hspace{-0.07in}\slash}

\begin{document}

\title{On the Question of a Possible Infrared Zero in the Beta Function of 
the Finite-$N$ Gross-Neveu Model}

\author{Gongjun Choi$^a$, Thomas Ryttov$^b$ and Robert Shrock$^a$}

\affiliation{(a) \ C. N. Yang Institute for Theoretical Physics and 
Department of Physics and Astronomy \\
Stony Brook University, Stony Brook, NY 11794, USA }

\affiliation{(b) \ The Center for Cosmology and Particle Physics Phenomenology CP$^3$-Origins, and the Danish Institute for Advanced Study DIAS,\\
University of Southern Denmark, Campusvej 55, DK-5230 Odense M, Denmark}

\begin{abstract}

We investigate whether the beta function of the finite-$N$ Gross-Neveu model, 
as calculated up to the four-loop level, exhibits  evidence for an infrared 
zero.  As part of our analysis, we calculate and analyze Pad\'e approximants
to this beta function and evaluate effects of scheme dependence.  From our
study, we find that in the range of coupling where the perturbative 
calculation of the four-loop beta function is reliable, it does not exhibit 
robust evidence for an infrared zero.  

\end{abstract}

\maketitle


\section{Introduction}
\label{intro}

The Gross-Neveu (GN) model \cite{gn} is a quantum field theory in $d=2$
spacetime dimensions with an $N$-component massless fermion $\psi_j$,
$j=1,...,N$, defined by the path integral
\beq
Z = \int \prod_x [{\cal D}\psi][{\cal D}\bar\psi] \, 
e^{i\int d^2x \, {\cal L}} \ , 
\label{z}
\eeq
with the Lagrangian density \cite{indices} 
\beq
{\cal L} = i\bar\psi \dslash \psi + \frac{g}{2} (\bar\psi\psi)^2 \ . 
\label{lag}
\eeq
This model is of interest because it exhibits, albeit in a lower-dimensional,
non-gauge-theory context, some properties of quantum chromodynamics (QCD),
namely asymptotic freedom, dynamical symmetry breaking of a certain chiral
symmetry, and the formation of a massive bound state of fermions.  These
properties were shown by an exact solution of the model in \cite{gn} in an $N
\to \infty$ limit that enabled Gross and Neveu to obtain nonperturbative
information about the theory. A semiclassical calculation
of the bound-state spectrum of the model was carried out in \cite{dhn}.

The Gross-Neveu model has also been studied at finite $N$, where it is not, in
general, exactly solvable.  In these studies, one again makes use of a property
that the model shares with QCD, namely asymptotic freedom, which allows one to
carry out reliable perturbative calculations at high Euclidean energy/momentum
scales $\mu$ in the deep ultraviolet (UV), where the running four-fermion
coupling, $g(\mu)$, approaches zero.  In this context, there is an interesting
and fundamental question: how does this running coupling $g(\mu)$ change as the
scale $\mu$ decreases from the deep UV to the infrared (IR) limit at $\mu = 0$?
This change of $g(\mu)$ as a function of $\mu$ is described by the
renormalization group (RG) \cite{rg} and the associated beta function, $\beta =
dg/dt$, where $dt = d\ln\mu$. The asymptotic freedom property is equivalent to
the fact that $\beta$ is negative in the vicinity of the origin, $g=0$, so that
this point is a UV fixed point (UVFP) of the renormalization group.  As $\mu$
decreases from the UV toward the IR, several different types of behavior of a
theory are, {\it a priori}, possible.  One is that the (perturbatively
calculated) beta function has no IR zero, so that as $\mu$ decreases, $g(\mu)$
eventually increases beyond the range where perturbative methods can be used to
study its RG evolution.  An alternative possibility is that $\beta$ has an IR
zero at sufficiently small coupling so that it can be studied using
perturbative methods.  An exact IR zero of $\beta$ would be an IR fixed point
(IRFP) of the renormalization group.  In the $N \to \infty$ limit used in
\cite{gn} to solve the model, the resultant beta function (given below in
Eq. (\ref{betagn})) does not exhibit any IR zero. Ref. \cite{schonfeld}
calculated $1/N$ corrections to the $N \to \infty$ limit in the Gross-Neveu
model and excluded the presence of an IR zero to this order.  However, to our
knowledge, there has not been an analysis of the beta function of the GN model
for finite $N$ to higher-loop order to address the question of whether it
exhibits evidence for an infrared fixed point.

In this paper we shall carry out this analysis of the beta function of the
finite-$N$ Gross-Neveu model to address and answer the question of whether this
function exhibits an IR zero.  We shall investigate the beta function to the
highest loop order to which it has been calculated, namely four loops, making
use of a recent computation of the four-loop term in Ref. \cite{gracey2016}.

This paper is organized as follows.  In Section \ref{background_section} we
review some background information about the Gross-Neveu model.  
In Section \ref{beta_section} we carry out
our analysis of the beta function of the finite-$N$ Gross-Neveu model up to the
four-loop level. In Section \ref{pade_section} we extend this analysis using
Pad\'e approximants.  Section \ref{scheme_section} contains an analysis of the
effect of scheme transformations on the beta function.  In Section
\ref{largeN_section} we comment further on the large-$N$ limit. 
Our conclusions are given in Section \ref{conc}.


\section{Some Relevant Background on the Gross-Neveu Model} 
\label{background_section}

Here we briefly review some relevant background concerning the Gross-Neveu
model. We first comment on some notation. In Ref. \cite{gn}, the coefficient in
front of the $(\bar\psi\psi)^2$ operator was written as a squared coupling,
which we denote as $(g_{GN}^2/2)$, while many subsequent works have written it
as $g/2$, so one has
\beq
g \equiv g_{GN}^2 \ . 
\label{ggn}
\eeq
The analysis of the model in \cite{gn} made use of a functional integral
identity to express the path integral as the $m \to \infty$ limit of a path
integral containing an auxiliary real scalar field $\phi$ with a mass $m$ and a
Yukawa interaction
\beq
{\cal L}_Y = g_{GN}m[\bar\psi \psi]\phi \ .
\label{yuk}
\eeq
Since $\phi$ is a real field, the hermiticity of ${\cal L}_Y$ implies that
$g_{GN}$ must be real, which, in conjunction with Eq. (\ref{ggn}), implies that
$g$ must be non-negative:
\beq
g \ge 0 \ .
\label{gpos}
\eeq

For $d=2$ (as more generally, for any even spacetime dimension), one can define
a product of Dirac gamma matrices, denoted $\gamma_5$, that satisfies the
anticommutation relation $\{\gamma_5,\gamma_\mu\}=0$ for all $\gamma_\mu$.
This $\gamma_5$ matrix also satisfies $\gamma_5^2=1$ and
$\gamma_5^\dagger=\gamma_5$.  (An explicit representation is $\gamma_0 =
\sigma_1$, $\gamma_1 = \sigma_2$, with $\gamma_0 \gamma_1 = i \gamma_5 =
i\sigma_3$, where $\sigma_j$ are the Pauli matrices.)  One can then define
chiral projection operators $P_{L,R} = (1/2)(1 \pm \gamma_5)$. 
As usual, one then defines left and right chiral components of the
fermion field as $\psi_L = P_L \psi$ and $\psi_R = P_R \psi$. 

The Gross-Neveu model is invariant under a discrete 
global ${\mathbb Z}_2$ group generated by the identity and the 
chiral transformation 
\beq
\psi \to \gamma_5 \psi \ . 
\label{psidiscrete}
\eeq
This discrete chiral transformation (\ref{psidiscrete}) takes $\bar\psi\psi \to
-\bar\psi\psi$, and hence this ${\mathbb Z}_2$ symmetry forbids (i) a mass term
in the Lagrangian (\ref{lag}) and (ii) the generation of a nonzero condensate
$\langle \bar\psi\psi\rangle$. This is true to all (finite) orders of
perturbation theory.

The Gross-Neveu model is also invariant under the continuous
global (cg) symmetry group
\beq
G_{cg} = {\rm U}(N) 
\label{cun}
\eeq
defined by the transformation
\beq
\psi \to U \psi \ , 
\label{psitran}
\eeq
where $U \in {\rm U}(N)$ (so $\bar\psi \to \bar\psi U^\dagger$). In terms of
the chiral components of the fermion field, the continuous global symmetry
transformation (\ref{psitran}) is $\psi_L \to U \psi_L$, $\psi_R \to U \psi_R$.
In contrast to the discrete $\gamma_5$ symmetry, the continuous symmetry
$G_{cg}$ leaves the operator $\bar\psi\psi$ invariant \cite{lsym}.

An exact solution of the theory was obtained in \cite{gn} in the limit
$N \to \infty$ and $g_{GN} \to 0$ with the product
\beq
\lambda \equiv g_{GN}^2 N \equiv gN
\label{lambda}
\eeq
a fixed and finite function of $\mu$.  We shall denote this as the LN limit
(i.e., the large-$N$ limit with the condition (\ref{lambda}) imposed).  In this
limit, there is a nonperturbative generation of a nonzero bilinear fermion
condensate, $\langle \bar \psi\psi \rangle$, dynamically breaking the discrete
${\mathbb Z}_2$ chiral symmetry.  In this limit, there is also the formation 
of a massive bound state of fermions.  

The beta function for $g_{GN}$ is
\beq
\beta_{GN} = \frac{dg_{GN}}{dt} \ , 
\label{betagndef}
\eeq
where $dt = d\ln\mu$. (The $\mu$ dependence of the coupling will often be
suppressed in the notation.)  This beta function is \cite{gn,otherbeta}
\beq
\beta_{GN} = -\frac{g_{GN}\lambda}{2\pi} \ . 
\label{betagn}
\eeq
The fact that this beta function is negative is an expression of the asymptotic
freedom of the theory.  This beta function does not exhibit any zero away from
the origin, i.e., any infrared zero.  However, since the calculation in
\cite{gn} was performed in the LN limit, this leaves open the possibility that
at finite $N$, there could be an IR zero in the beta function that would
disappear in the LN limit. We discuss this LN limit further in Section
\ref{largeN_section} below.


\section{Beta Function for General $N$}
\label{beta_section}

Although the Gross-Neveu model is not, in general, solvable away from the LN
limit, there has also been interest over the years in analyzing it for finite
$N$.  In terms of the coupling $g$, the beta function of the finite-$N$ GN
model is
\beq
\beta = \frac{dg}{dt} \ , 
\label{betag}
\eeq
where, as before, $dt = d\ln\mu$.  For our purposes, it will be convenient to
introduce a variable $a$ that includes the factor $1/(2\pi)$ resulting from 
Feynman integrals in $d=2$ dimensions, namely 
\beq
a = \frac{g}{2\pi} = \frac{g_{GN}^2}{2\pi} \ . 
\label{adef}
\eeq
The model defined by the Lagrangian of Eq. (\ref{lag}) can be generalized with
the addition of further four-fermion operators \cite{gn,rossi89}. The
regularization and renormalization of the Gross-Neveu model has been carried
out in this more general context \cite{rossi89}-\cite{rossi91}, 
\cite{gracey2016}.

As was true of other theories, such as the nonlinear $\sigma$ model
\cite{nlsm}, one may consider this model in spacetime dimension $d > 2$.  At
finite $N$, the model is not renormalizable for $d > 2$, since the Maxwellian
dimension of a four-fermion operator is $2(d-1)$, which is larger than $d$ if
$d > 2$. As in the case of the nonlinear $\sigma$ model \cite{nlsm}, in the $N
\to \infty$ limit, one can still solve the model and study its properties.
Alternatively, for finite $N$, one can regard it as a low-energy effective
field theory.  With this generalization and $d \gsim 2$, $\beta$ has the form
\beqs
\beta & = & g\Big [ d-2 + \sum_{\ell=1}^\infty b_\ell \, \Big ( \frac{g}{2\pi}
\Big )^\ell \ \Big ] \cr\cr
& = & 2\pi a\Big [d-2 + \sum_{\ell=1}^\infty b_\ell \, a^\ell \ \Big ] \ , 
\label{betaseries}
\eeqs
where $b_\ell a^\ell$ is the $\ell$-loop term.  The $n$-loop ($n\ell$) beta
function, denoted $\beta_{n\ell}$, is obtained by the replacement of
$\ell=\infty$ by $\ell=n$ in Eq. (\ref{betaseries}). Early discussions of the
GN model for $d > 2$ include \cite{gn} and \cite{vasiliev}; for more recent
work see, e.g., \cite{gracey2016}, \cite{manashov}, and, for condensed-matter
applications, \cite{cm}, and references therein.  In this paper, aside from
some comments in Section \ref{largeN_section}, we will restrict ourselves to
the Gross-Neveu model in $d=2$, where $g$ is dimensionless.

The $\ell=1$ and $\ell=2$ loop terms in $\beta$ are independent of the scheme
used for regularization and renormalization, while the terms at loop order
$\ell \ge 3$ are scheme-dependent.  The beta function was calculated up to
two-loop level in \cite{wetzel85}, with the results
\beq
b_1 = -2(N-1)
\label{b1}
\eeq
and
\beq
b_2 = 2(N-1) \ . 
\label{b2}
\eeq
(See also \cite{destri} for a two-loop calculation in a related Thirring 
model.) The fact that $b_1$ in Eq. (\ref{b1}) is negative means that in $d=2$,
this theory is asymptotically free for any finite $N > 1$ as well as in the $N
\to \infty$ limit considered in \cite{gn}.

The three-loop coefficient, $b_3$, was
calculated in \cite{gracey90,rossi91} in the commonly used scheme with
dimensional regularization and modified minimal subtraction, denoted
$\overline{\rm MS}$ \cite{msbar}, yielding the result
\beq
b_3 = \frac{(N-1)(2N-7)}{2} \ . 
\label{b3}
\eeq
Recently, the four-loop coefficient, $b_4$ has been calculated, again in the 
$\overline{\rm MS}$ scheme, to be \cite{gracey2016} 
\beq
b_4 = \frac{1}{3}(N-1) \Big [ -2N^2 - 19N + 24 - 6(11N-17)\zeta_3 \Big ] \ , 
\label{b4}
\eeq
where $\zeta_s = \sum_{n=1}^\infty n^{-s}$ is the Riemann zeta function. 

We comment on the dependence of the beta function coefficients on $N$.  The
property that these coefficients all contain a factor of $(N-1)$ is a
consequence of the fact that for $N=1$ the GN model is equivalent to the
massless abelian Thirring model \cite{thirring}, which has an identically zero
beta function \cite{klaiber,lowenstein}. Note that this statement about the
beta function of the Thirring model is scheme-independent; if a beta function
vanishes in one scheme, then it vanishes in all other schemes reached by
acceptable (nonsingular) scheme transformations \cite{sch}. It follows that all
of the coefficients $b_\ell$ contain a factor of $(N-1)$. Therefore, it is only
necessary to analyze the beta function of the Gross-Neveu model for $N > 1$,
where it is nonvanishing, and we will thus restrict to the physical integral
values $N \ge 2$ henceforth.  We next discuss how the $b_\ell$ depend on $N$ in
the relevant range $N > 1$.  For this discussion, we consider $N$ to be
extended from the positive integers to the real numbers.  The three-loop
coefficient $b_3$ is a monotonically increasing function of $N$ that is
negative for $N < 7/2$, vanishes for $N=7/2$, and is positive for $N > 7/2$.
Thus, for physical, integral values, $b_3 < 0$ if $N=2$ or $N=3$ and $b_3 > 0$
if $N \ge 4$.  The coefficient $b_4$ is negative for large $N$ and is positive
for $N$ in the interval
\beq
N_{b4z,m} < N < N_{b4z,p} \ ,
\label{ninterval}
\eeq
where the subscript $b4z$ stands for ``$b_4$ zero'' and
\beq
N_{b4z,(p,m)} = \frac{-19-66\zeta_3 \pm
  \sqrt{553+3324\zeta_3+4356\zeta_3^2}}{4}
\label{Nb4z}
\eeq
with $(p,m)$ corresponding to the $\pm$ sign.  These have the values
$N_{b4z,m} = -50.616$ and $N_{b4z,p}=1.448$ to the given floating-point
accuracy. Thus, in the relevant range $N > 1$ under consideration here, $b_4$
is negative.

We proceed to investigate the question of whether the beta function for the
Gross-Neveu model at finite $N$ exhibits evidence for an infrared 
zero. We denote an IR zero of the $n$-loop beta function $\beta_{n\ell}$
as $a_{IR,n\ell}$, and the corresponding value of $g$ as $g_{IR,n\ell}=2\pi
a_{IR,n\ell}$.  This IR zero of beta is a zero for positive $a$ closest to the
origin (if there is such a zero), which one would thus reach as $\mu$
decreases from the deep UV at large $\mu$ to the IR at small $\mu$ and $a$
increases from 0.  At the two-loop level, $\beta_{2\ell}$ has an IR zero at
\beq
a_{IR,2\ell} = -\frac{b_1}{b_2} = 1 \ , 
\label{air_2loop}
\eeq
i.e., $g_{IR,2\ell}=2\pi$.  Note that this value is independent of $N$. To
judge whether this constitutes convincing evidence of an IR zero in the
beta function, it is necessary to determine if higher-loop calculations confirm
it. We next carry out this task. 

At the three-loop level, the condition that $\beta_{3\ell}=0$ away from the
origin is the quadratic equation $b_1+b_2a+b_3 a^2=0$.  This has two solutions,
\beq
a=\frac{2[-1 \pm \sqrt{2(N-3)} \ ]}{2N-7} \ .
\label{asol3loop}
\eeq
If $N < 3$, then these solutions are complex and
hence unphysical.  If $N=3$, these roots coincide, so that $a_{IR,3\ell}=2$,
i.e., $g_{IR,3\ell}=4\pi$.  For $N \ge 3$, there is only one physical root,
namely 
\beq
a_{IR,3\ell} = \frac{2[-1+\sqrt{2(N-3)} \ ]}{2N-7} \ . 
\label{air_3loop}
\eeq
However, this is not, in general, close to the two-loop zero of the beta
function at $a_{IR,2\ell}=1$.   Furthermore, while $a_{IR,2\ell}=1$ is
independent of $N$, $a_{IR,3\ell}$ has a completely different behavior as a
function of $N$; it decreases monotonically with $N$ in the 
interval $N \ge 3$ over which it is physical and approaches zero asymptotically
like 
\beq
a_{IR,3\ell} \sim \sqrt \frac{2}{N} - \frac{1}{N} + 
O \Big ( \frac{1}{N^{3/2}} \Big ) \quad {\rm as} \ N \to \infty \ . 
\label{air_3loop_largeN}
\eeq

At the four-loop level, the condition that $\beta_{4\ell}=0$ away from the
origin is the cubic equation
\beq
b_1+b_2a+b_3 a^2 + b_4 a^3=0 \ . 
\label{acubic}
\eeq
The nature of the roots of this equation is determined by the discriminant,
\beq
\Delta_3 = b_2^2b_3^2-27b_1^2b_4^2 - 4(b_1b_3^3+b_4b_2^3) + 18b_1b_2b_3b_4 \ .
\label{disc3}
\eeq
This discriminant is negative for the relevant range $N \ge 2$ (indeed, it is
negative for all real $N$).  This implies that Eq. (\ref{acubic}) has one real
root and a pair of complex-conjugate roots.  The real root is negative and
hence is unphysical, since it violates the positivity requirement
(\ref{gpos}). Moreover, since it is negative, it is clearly incompatible with
the values of $a_{IR,2\ell}$ and $a_{IR,3\ell}$, which are positive (discarding
the unphysical complex value of $a_{IR,3\ell}$ at $N=2$).  We therefore do not
label this root as $a_{IR,4\ell}$, but instead as $a_{rt,4\ell}$, where $rt$
stands simply for the real root of Eq. (\ref{acubic}).  We find that the
magnitude of $a_{rt,4\ell}$ decreases toward zero monotonically as $N$
increases in the relevant interval $N \ge 2$, with the asymptotic behavior
\beq
a_{rt,4\ell} \sim -\frac{3^{1/3}}{N^{2/3}} + \frac{1}{2N} + 
O \Big ( \frac{1}{N^{4/3}} \Big ) \quad {\rm as} \ N \to \infty \ . 
\label{art_largeN}
\eeq
We list the values of $a_{IR,2\ell}$, $a_{IR,3\ell}$, and
$a_{rt,4\ell}$ in Table \ref{air_nloop_values} for $N$ from 2 to 10 and for
three representative larger values, $N=100$, 300, and $10^3$.

\begin{table}
\caption{\footnotesize{Values of $a_{IR,2\ell}$, $a_{IR,3\ell}$, and
$a_{rt,4\ell}$ for the beta function of the Gross-Neveu model, as a
function of $N$.  Here, the three-loop and four-loop coefficients $b_3$ and 
$b_4$ are calculated in the $\overline{\rm MS}$ scheme. If $N=2$, then 
the zeros of $\beta_{3\ell}$ at nonzero $a$ form an unphysical complex (cmplx)
pair.  As indicated, all of the values of $a_{rt,4\ell}$ are negative and hence
unphysical. See text for further details.}}

\begin{center}
\begin{tabular}{|c|c|c|c|} \hline\hline
$N$ & $a_{IR,2\ell}$ & $a_{IR,3\ell}$ & $a_{rt,4\ell}$ \\
\hline
2      &  1  &  cmplx  & $-0.573$   \\
3      &  1  &  2.000  & $-0.370$   \\
4      &  1  &  0.828  & $-0.302$   \\
5      &  1  &  0.667  & $-0.264$   \\
6      &  1  &  0.580  & $-0.239$   \\
7      &  1  &  0.522  & $-0.220$   \\
8      &  1  &  0.481  & $-0.205$   \\
9      &  1  &  0.448  & $-0.194$   \\
10     &  1  &  0.422  & $-0.184$   \\
100    &  1  &  0.134  & $-0.0567$  \\
300    &  1  &  0.0788 & $-0.0295$  \\
$10^3$ &  1  &  0.0438 & $-0.0138$  \\
\hline\hline
\end{tabular}
\end{center}
\label{air_nloop_values}
\end{table}

In our discussion above, we had stated that in order to judge whether the
result for $a_{IR,2\ell}$ constitutes convincing evidence of an IR zero in the
beta function, it is necessary to determine if higher-loop calculations confirm
it.  A necessary condition for the reliability of a perturbative calculation is
that if one calculates some quantity to a given loop order, then there should
not be a large fractional change in this quantity if one computes it to one
higher order in the loop expansion. This condition applies, in particular, to
the calculation of a putative zero of the beta function. Quantitatively, in
order for the perturbative calculation of the IR zero of a beta function to be
reliable, it is necessary that the fractional difference 
\beq
\frac{|a_{IR,(n-1)\ell} - a_{IR,n\ell}|}
{\frac{1}{2}[a_{IR,(n-1)\ell}+ a_{IR,n\ell}]}
\label{fracdif}
\eeq
should be reasonably small and should tend to decrease with increasing loop
order, $n$.  As is evident both from our analytic formulas and from the
numerical results listed in Table \ref{air_nloop_values}, this 
necessary condition is not satisfied in the present case.

\begin{figure}
  \begin{center}
    \includegraphics[height=8cm,width=6cm]{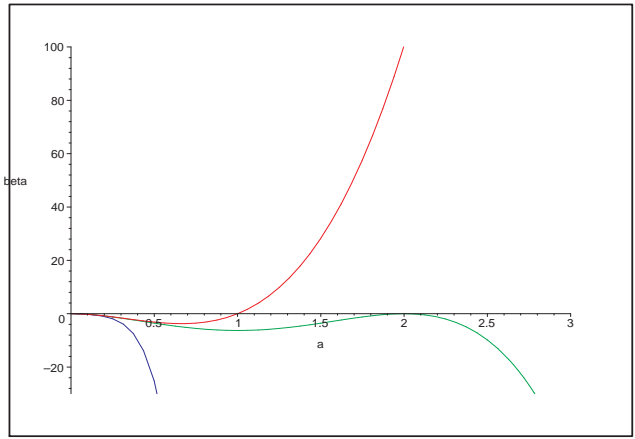}
  \end{center}
\caption{\footnotesize{Plot of the $n$-loop $\beta$ function 
$\beta_{a,n\ell}$ of the Gross-Neveu model as a function of $a$ for $N=3$ and
(i) $n=2$ (red), (ii) $n=3$ (green), and (iii) $n=4$ (blue) (colors in 
online version). At $a=0.16$, going from bottom to top, the curves are 
$\beta_{4\ell}$, $\beta_{2\ell}$, and $\beta_{3\ell}$.}}
\label{beta_N3}
\end{figure}

\begin{figure}
  \begin{center}
    \includegraphics[height=8cm,width=6cm]{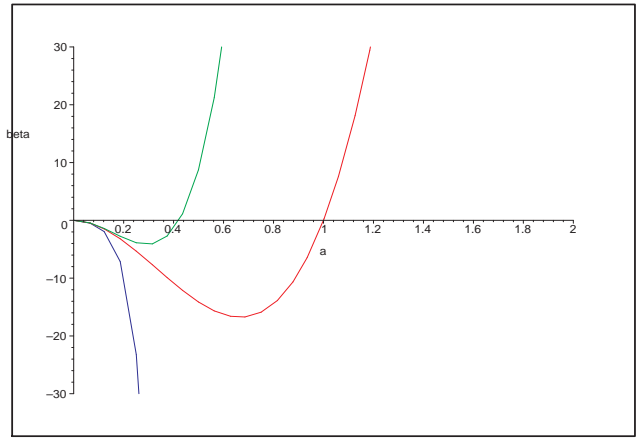}
  \end{center}
\caption{\footnotesize{Plot of the $n$-loop $\beta$ function 
$\beta_{a,n\ell}$ of the Gross-Neveu model as a function of $a$ for $N=10$ and
(i) $n=2$ (red), (ii) $n=3$ (green), and (iii) $n=4$ (blue) (colors in 
online version). At $a=0.2$, going from bottom to top, the curves are 
$\beta_{4\ell}$, $\beta_{2\ell}$, and $\beta_{3\ell}$.}}
\label{beta_N10}
\end{figure}

The reason for this is clear from a plot of the beta functions $\beta_{n\ell}$
at loop orders $n=2$, $n=3$, and $n=4$.  This shows that the IR zero in the
two-loop beta function occurs at a value of $a$ that is too large for the
perturbative calculation to be reliable.  In Figs. \ref{beta_N3} and
\ref{beta_N10} we plot the two-loop, three-loop, and four-loop beta functions
for the Gross-Neveu model as functions of $a$ for two illustrative values of
$N$, namely $N=3$ and $N=10$.  As is evident from these plots, the beta
function does not satisfy the necessary criterion for the reliability of a
calculation of an IR zero.  For the IR zero of the two-loop beta function at
$a_{IR,2\ell}=1$ to be reliable, one requires that the curves for the
three-loop and four-loop beta functions should agree approximately with the
curve for the two-loop beta function for $a \simeq 1$, and that these
higher-loop beta functions should thus have respective IR zeros that are close
to the two-loop zero at $a_{IR,2\ell}=1$.  But this is not the case; for $N=3$,
$\beta_{3\ell}$ has a double zero at the larger value, $a_{IR,3\ell}=2$ and
then goes negative again, while $\beta_{4\ell}$ has no IR zero in the physical
region, $a > 0$.  For $N=10$ the three-loop beta function $\beta_{3\ell}$
vanishes at a smaller value of $a$ than $a=1$ (and this value, $a_{IR,3\ell}$
decreases as $N$ increases), while the four-loop beta function $\beta_{4\ell}$
again has no IR zero in the physical region, $a > 0$.  The behavior illustrated
for $N=10$ is generic for other values of $N \ge 4$.  Indeed, the curves for
these beta functions at loop order $n=2, \ 3, \ 4$ only agree with each
other close to the origin, and deviate strongly from each other before one gets
to values of $a$ where a zero occurs.  Specifically, for $N=3$, $\beta_{2\ell}$
and $\beta_{3\ell}$ only agree with each other for $a$ up to about 0.5, while
$\beta_{4\ell}$ deviates from these lower-loop beta functions as $a$ increases
beyond approximately 0.2. As $N$ increases, these deviations occur for smaller
$a$.  Thus, for $N=10$, $\beta_{2\ell}$ and $\beta_{3\ell}$ only agree with
each other for $a$ up to roughly 0.15, while $\beta_{4\ell}$ deviates from
these lower-loop beta functions as $a$ increases beyond about 0.08.


These results are similar to what was found in a search for a UV zero in the
beta function of an IR-free theory, namely the O($N$) $\lambda|{\vec \phi}|^4$
scalar field theory in $d=4$ spacetime dimensions \cite{lam}. In that theory,
although the two-loop beta function exhibits a UV zero, higher-loop
calculations up to five-loop order for general $N$ and up to six-loop order for
$N=1$ do not confirm the two-loop result, and the reason was found to be that
the two-loop UV zero occurs at too large a value of the quartic coupling for
the two-loop perturbative calculation to be applicable and reliable. 


\section{Analysis with Pad\'e Approximants}
\label{pade_section} 

In this section we carry out a further investigation of a possible IR fixed
point in the renormalization-group flow for the Gross-Neveu model by
calculating and analyzing Pad\'e approximants (PAs) to the beta function at
three-loop and four-loop level.  Since we are interested in a possible zero of
the beta function away from the origin, it will be convenient to deal with a
reduced ($rd$) beta function,
\beq
\beta_{rd} \equiv \frac{\beta}{2\pi b_1 a^2} = 
1 + \frac{1}{b_1} \sum_{\ell=2}^\infty
b_\ell a^{\ell-1} \ . 
\label{betareduced}
\eeq
The $n$-loop reduced beta function with $n \ge 2$, denoted $\beta_{rd,n\ell}$,
is obtained from Eq. (\ref{betareduced}) by replacing $\ell=\infty$ by $\ell=n$
as the upper limit in the summand.  This $n$-loop reduced beta function is thus
a polynomial of degree $n-1$ in $a$. The $[p,q]$ Pad\'e approximant to this
polynomial is the rational function
\beq
[p,q]_{\beta_{rd,n\ell}} =
\frac{1+\sum_{j=1}^p \, n_j x^j}{1+\sum_{k=1}^q d_k \, x^k}
\label{pqx}
\eeq
with
\beq
p+q=n-1 \ ,
\label{pqn}
\eeq
where the $n_j$ and $d_k$ are $a$-independent coefficients of the respective
polynomials in the numerator and denominator of
$[p,q]_{\beta_{rd,n\ell}}$. (Our notation follows \cite{smpade}.) 
Hence, at a given $n$-loop order, there are $n$
Pad\'e approximants that one can calculate, namely 
\beq
\{ \ [n-k,k-1]_{\beta_{rd,n\ell}} \ \} \quad {\rm with} \
1 \le k \le n \ .
\label{padeset}
\eeq
These provide rational-function approximations of the series expansion for 
$\beta_{rd,n\ell}$ that fits this series to the loop order $n$.  As in our
earlier work, e.g., \cite{bvh,flir}, these provide an
alternate approach to investigating zeros of a beta function. 

We shall label one of the $p$ zeros of a
$[p,q]_{\beta_{rd,n\ell}}$ Pad\'e
approximant as $[p,q]_{zero}$ and one of the $q$ poles of this
approximant as $[p,q]_{pole}$; in each case, the value of $n$ is given by
Eq. (\ref{pqn}) as $n=p+q+1$.  At the $n$-loop level, the Pad\'e approximant
$[n-1,0]_{\beta_{rd,n\ell}}$ is equal to the reduced $n$-loop beta function 
$\beta_{rd,n\ell}$ itself, which we have already analyzed in the previous
section, and the PA $[0,n-1]_{\beta_{rd,n\ell}}$ has no zeros, and hence is not
useful for our study.  Hence, at the $n$-loop level, we focus on the $n-2$ 
PAs 
$[p,q]_{\beta_{rd,n\ell}}$ with $[p,q]=[n-k,k-1]$ having $2 \le k \le n-1$. 

At the $n=3$ loop level, we thus consider the $[1,1]_{\beta_{rd,3\ell}}$
Pad\'e approximant. This is 
\beq
[1,1]_{\beta_{rd,3\ell}}=\frac{1+\Big (\frac{b_2}{b_1}-\frac{b_3}{b_2} \Big )a}
{1-\Big ( \frac{b_3}{b_2}\Big ) a} = \frac{1- \Big ( \frac{2N-3}{4} \Big )a}
           {1- \Big ( \frac{2N-7}{4} \Big )a} \ . 
\label{pade11}
\eeq
where the coefficients $b_1$, $b_2$, and $b_3$ were given in
Eqs. (\ref{b1})-(\ref{b3}) above. This [1,1] PA has a zero at
\beq
[1,1]_{zero} = \frac{4}{2N-3}
\label{p11zero}
\eeq
and a pole at
\beq
[1,1]_{pole} = \frac{4}{2N-7} \ . 
\label{p11pole}
\eeq
The $a=[1,1]_{pole}$ is not relevant, since if $N = 2$ or 3, it has the
respective negative and hence unphysical values $-4/3$ and $-4$, while for 
$N \ge 4$, it lies farther from the origin than
the zero.  This is clear from the fact that the difference 
\beq
[1,1]_{pole} -[1,1]_{zero} = \frac{16}{(2N-3)(2N-7)}
\label{p11polezerodif}
\eeq
is positive for this range $N \ge 4$.  Since the  $[1,1]_{pole}$ lies farther
from the origin than $[1,1]_{zero}$, the coupling $a=a(\mu)$ never reaches the
pole as $\mu$ decreases from large values in the UV to $\mu=0$ and thus
$a(\mu)$ increases from 0 to $[1,1]_{zero}$.  We list the values of the zero of
the $[1,1]_{\beta_{rd,3\ell}}$ Pad\'e approximant in Table \ref{pades}.  For $N
\ge 3$, the value of $a=[1,1]_{zero}$ is smaller than $a_{IR,3\ell}$ and
decreases more rapidly to zero as $N \to \infty$ than $a_{IR,3\ell}$.  If
$N=3$, the comparison cannot be made, since $a_{IR,3\ell}$ is complex. Thus,
this analysis of the [1,1] Pad\'e approximant to the reduced three-loop beta
function, $\beta_{rd,3\ell}$ yields further evidence against a (reliably
calculable) IR zero in the beta function up to the three-loop level.

At the $n=4$ loop level, there are two Pad\'e approximants to analyze, 
namely $[2,1]_{\beta_{rd,4\ell}}$ and $[1,2]_{\beta_{rd,4\ell}}$. We calculate
\beq
[2,1]_{\beta_{rd,4\ell}}=\frac{1+\Big ( \frac{b_2}{b_1}-\frac{b_4}{b_3}\Big )a
+ \Big ( \frac{b_3}{b_1}-\frac{b_2b_4}{b_1b_3} \Big )a^2}
{1- \frac{b_4}{b_3}a} \ , 
\label{pade21}
\eeq
where the coefficients $b_n$ were given in Eqs. (\ref{b1})-(\ref{b4}).  The
zeros of the numerator occur at $a=[2,1]_{zero,(i,ii)}$, where 
\beqs
& & [2,1]_{zero,(i,ii)} = \cr\cr
& &  \frac{b_2b_3-b_1b_4 \pm 
\Big [ b_1^2b_4^2+b_2^2b_3^2 -4b_1b_3^3+2b_1b_2b_3b_4 \Big ]^{1/2}}
{2(b_2b_4-b_3^2)} \ . \cr\cr
& & 
\label{pade21zeros}
\eeqs
and the subscripts $i$ and $ii$ correspond to the $\pm$ sign in front of the
square root.  It is straightforward to substitute the explicit expressions for
the coefficients $b_2$, $b_3$, and $b_4$ in Eq. (\ref{pade21zeros}), but the
resultant expressions for these quadratic roots in terms of the explicit
coefficients $b_n$, $1 \le n \le 4$ are somewhat lengthy, so we do not display
them.  The pole of the $[2,1]_{\beta_{rd,4\ell}}$ PA occurs at $a=
[2,1]_{pole}$, where
\beqs
& & [2,1]_{pole} = \frac{b_3}{b_4} \cr\cr
& = & -\frac{3(2N-7)}{2[2N^2+19N-24+6(11N-17)\zeta_3]} \ . 
\label{p21pole}
\eeqs
If one has a series expansion of a function that contains $n_{zero}$ zeros 
and $n_{pole}$ poles, and one calculates $[r,s]$ Pad\'e approximants to this 
series with $r > n_{zeros}$ and $s > n_{poles}$, the approximants typically
exhibit sets of nearly coincident zero-pole pairs in addition to 
fitting the actual zeros and poles of the function 
(e.g., see \cite{smpade,flir}).  These nearly coincident zero-pole pairs may
thus be ignored.  This happens in the present case.  For example, for $N=3$,
the $[2,1]_{\beta_{rd,4\ell}}$ PA has a zero at $a=0.99773$, a zero at
$a=0.009015$ and a pole at $a=0.009015$, and similarly for other values of
$N$.  In Table \ref{pades} we list the first zero, denoted $[2,1]_{zero,i}$, 
as a function of $N$.

\begin{table}
\caption{\footnotesize{Values of $[1,1]_{zero}$ from [1,1] Pad\'e approximant
    to the reduced three-loop beta function, $\beta_{rd,3\ell}$, and 
$[2,1]_{zero,i}$ from the [2,1] Pad\'e approximant to the four-loop beta
function, $\beta_{rd,4\ell}$.  See text for further details. }}
\begin{center}
\begin{tabular}{|c|c|c|c|} \hline\hline
$N$ & $[1,1]_{zero}$ & $[2,1]_{zero,i}$ \\
\hline
2     &  4.000   &  0.940    \\ 
3     &  1.333   &  0.998    \\
4     &  0.800   &  0.999    \\
5     &  0.571   &  0.992    \\
6     &  0.444   &  0.982    \\
7     &  0.364   &  0.9725   \\
8     &  0.308   &  0.963    \\
9     &  0.267   &  0.953    \\
10    &  0.235   &  0.943    \\
100   &  0.0203  &  0.683    \\
300   &  0.00670 &  0.615    \\
$10^3$&  0.00200 &  0.585    \\
\hline\hline
\end{tabular}
\end{center}
\label{pades}
\end{table}

We calculate the $[1,2]_{\beta_{rd,4\ell}}$ Pad\'e approximant to be 
\beq
[1,2]_{\beta_{rd,4\ell}}=\frac{1+
\Big [\frac{b_1^2b_4+b_2^3-2b_1b_2b_3}{b_1(b_2^2-b_1b_3)} \Big ]a}
{1 + \Big ( \frac{b_1b_4-b_2b_3}{b_2^2-b_1b_3} \Big ) a 
+ \Big ( \frac{b_3^2-b_2b_4}{b_2^2-b_1b_3} \Big ) a^2 } \ . 
\label{pade12}
\eeq
The two poles of the $[1,2]_{\beta_{rd,4\ell}}$ approximant occur at
$a=[1,2]_{pole,(i,ii)}$, where 
\begin{widetext}
\beq
[1,2]_{pole,(i,ii)}= \frac{b_1b_4 - b_2b_3 \pm 
\Big [ b_1^2b_4^2-3b_2^2b_3^2+4b_1b_3^3+4b_2^3b_4-6b_1b_2b_3b_4 \Big ]^{1/2}}
{2(b_2b_4-b_3^2)} \ . 
\label{p12poles}
\eeq
\end{widetext}
The zero of this approximant occurs at $a=[1,2]_{zero}$, where 
\beqs
& & [1,2]_{zero} = \frac{b_1(b_1 b_3-b_2^2)}{b_1^2b_4+b_2^3-2b_1b_2b_3} \cr\cr
& = & -\frac{3(2N-3)}{2[2N^2+13N-9+6(11N-17)\zeta_3]} \ . 
\label{p12zero}
\eeqs
Both of the poles $[1,2]_{pole,i}$ and $[1,2]_{pole,ii}$ are negative.
Furthermore, we find that this approximant has nearly coincident zero-pole
pairs, which thus can both be ignored.  For example, for $N=3$, the zero occurs
at $a=-0.027540$ while one of the poles occurs at the nearly equal value,
$a=-0.027556$, and the other pole is at $a=-0.97919$. Similar results hold for
other values of $N$, i.e., the $[1,2]_{\beta_{rd,4\ell}}$ PA has a nearly
coincident zero-pole pair (at negative $a$) together with a second unphysical
pole at negative $a$.

As we have discussed, the four-loop beta function yields a negative real root,
in strong disagreement with the two-loop and three-loop beta functions.  At
this four-loop level, the [1,2] PA does not exhibit any true zero, but only a
zero that is nearly coincident with a pole and hence can be identified as an
artifact.  The [2,1] PA yields a zero, but it is at a completely different
value than the only real root of the actual four-loop beta function,
$a_{rt,4\ell}$. Thus, our analysis of the [2,1] and [1,2] Pad\'e approximants
to the four-loop (reduced) beta function yield further evidence against a
robust IR zero in this four-loop beta function.


\section{Analysis Using Scheme Transformations}
\label{scheme_section} 

Since the coefficients $b_\ell$ with $\ell \ge 3$ in the beta function are
scheme-dependent, it is necessary to check that the conclusions from our
analysis of the beta function with $b_3$ and $b_4$ calculated in the
$\overline{\rm MS}$ scheme are robust with respect to scheme
transformations. To begin, we study scheme transformations that are designed to
remove higher-loop terms in the beta function. We first review some relevant
background.  In \cite{sch}, formulas were derived for the coefficients
$b_\ell'$ resulting from a general scheme transformation $f(a')$ of the form
\beq
a = a'f(a') \ . 
\label{aap}
\eeq
Since a scheme transformation has no effect in the case of a free field theory,
$f(a')$ satisfies the condition that $f(0)=1$.  Expressing $f(a')$ as a power 
series in $a'$, one has
\beq
f(a') = 1 + \sum_{s=1}^{s_{max}} k_s (a')^s \ ,
\label{faprime}
\eeq
where the $k_s$ are constants and $s_{max}$ may be finite or infinite. It
follows that the Jacobian of this transformation, $J=da/da'$ satisfies the
condition $J(0)=1$ and has the expansion 
\beq
J = 1 + \sum_{s=1}^{s_{max}} (s+1)k_s(a')^s \ . 
\label{j}
\eeq
Then in the transformed scheme, the coefficients of the three-loop and
four-loop terms in the beta function are \cite{sch}
\beq
b_3' = b_3 + k_1b_2+(k_1^2-k_2)b_1 \ , 
\label{b3prime}
\eeq
\beq
b_4' = b_4 + 2k_1b_3+k_1^2b_2+(-2k_1^3+4k_1k_2-2k_3)b_1 \ , 
\label{b4prime}
\eeq
and so forth for higher $b'_\ell$. 

In \cite{sch} a set of conditions was given that should be obeyed by a
nonpathological scheme transformation. Condition C$_1$ was that the scheme
transformation must map a physical (real, positive) $a$ to a real positive
$a'$, since a map that yields a negative or complex value of $a'$ would violate
the unitarity of the theory. As condition C$_2$, we required that the scheme
transformation should preserve perturbativity, and hence should not map a small
or moderate value of $a$ to an excessively large value of $a'$ or vice versa.
Condition C$_3$ stated that the Jacobian $J$ should not vanish or diverge,
since otherwise the transformation would be singular.  More generally, if $J$
were to become too small or too large, it could lead to a violation of
condition C$_2$.  Finally, condition C$_4$ was that if a beta function
exhibited a zero at a sufficiently small value as to be perturbatively
reliable, then a scheme transformation should not alter this property.
Ref. \cite{sch} also gave the first explicit scheme transformation to set
$b_\ell'=0$ for $\ell \ge 3$, at least in the local vicinity of the origin, but
it also showed that this does not, in general, work to remove these higher-loop
terms at a point located away from the origin, i.e., an IR zero in an
asymptotically free theory or a UV zero in an IR-free theory. The reason, as
shown in \cite{sch} and \cite{sch23}, if one attempts to apply such a scheme
transformation to remove these higher-loop terms at a point away from the
origin, then the transformation violates one or more of the conditions
C$_1$-C$_4$ for acceptability. As in \cite{sch23}, we denote the scheme
transformation presented in \cite{sch} (with $s_{max}=m$) that removes the
coefficients in the beta function up to loop order $\ell=m+1$, at least near
the origin, as $S_{R,m}$.

We proceed with our analysis with the $S_{R,m}$ scheme transformation. 
The $S_{R,2}$ transformation has \cite{sch} 
\beq
k_2 = \frac{b_3}{b_1}
\label{k2}
\eeq
and the $S_{R,3}$ transformation has this $k_2$ and 
\beq
k_3 = \frac{b_4}{2b_1} \ . 
\label{k3}
\eeq
We begin by determining whether the scheme transformation $S_{R,2}$ can be
applied in the relevant region of $a$ where we need to apply it to set $b_3'=0$
and thus remove the three-loop term in the beta function.  Since the
(scheme-independent) two-loop value is $a_{IR,2\ell}=a_{IR,2\ell}' = 1$, the
relevant region is in the neighborhood of $a=1$.  This $S_{R,2}$ transformation
is defined by Eq. (\ref{faprime}) with $s_{max}=2$ and $k_2$ given by
Eq. (\ref{k2}).  If the application of this $S_{R,2}$ transformation in the
vicinity of $a=$ were possible, then it would follow from Eq. (\ref{b4prime})
that $b_4'=b_4$.  For $S_{R,2}$, Eq. (\ref{aap}) is
\beq
S_{R,2} \ \Longrightarrow \  a=a'[1+k_2 (a')^2] = a'\Big
[1+\frac{b_3}{b_1}(a')^2 \Big ] \ . 
\label{aeq_sr2}
\eeq
Solving Eq. (\ref{aeq_sr2}) for $a'$, we obtain three roots, and we require
that at least one of these should be a physical (real, positive) value for $a$
in the relevant range of values comparable to $a_{IR,2\ell}=1$.  We find that
this necessary condition, C$_1$, is not satisfied. Instead, two of the
solutions of Eq. (\ref{aeq_sr2}) for $a'$ form a complex-conjugate pair, while
the third is negative.  For example, for $a=a_{IR,2\ell}=1$ and $N=4$, the
three solutions for $a'$ are $1.191 \pm 0.509i$ and $-2.383$, while for $N=10$,
the three solutions for $a'$ are $0.4125 \pm 0.450i$ and $-0.825$.  The
Jacobian also exhibits pathological behavior; $J$ is given by
\beqs
S_{R,2} \ \Longrightarrow \  J & = & 1 + 3k_2(a')^2 
                                 =   1 + \frac{3b_3}{b_1}(a')^2 \cr\cr
& = & 1 - \frac{3(2N-7)}{4} \, (a')^2 \ . 
\label{j_sr2}
\eeqs
For $a_{IR,2\ell}=a_{IR,2\ell}'=1$, $J=(25-6N)/4$, which decreases through zero
as $N$ (continued to the real numbers) increases through the value $N=25/6$,
violating condition C$_3$.  It is therefore not
possible to use this scheme transformation to remove the three-loop term in the
beta function in the region of $a$ where we are trying to do this, namely the
neighborhood of the (scheme-independent) value $a=a_{IR,2\ell}=1$.

We can also investigate whether the scheme transformation $S_{R,3}$ is
physically acceptable to be applied in the relevant range of values of $a$,
namely $a=a_{IR,2\ell}=1$. This transformation is defined by
Eq. (\ref{faprime}) with $s_{max}=3$ and $k_2$ and $k_3$ given by Eqs. 
(\ref{k2}) and (\ref{k3}):
\beqs
S_{R,3} \ \Longrightarrow \ a & = & a'[1+k_2 (a')^2+k_3(a')^3] \cr\cr
& = & a'\Big [1+\frac{b_3}{b_1}(a')^2 + \frac{b_4}{2b_1} (a')^3 \Big ] \ . 
\label{aeq_sr3}
\eeqs
The Jacobian for this transformation is
\beqs
S_{R,3} \ \Longrightarrow \  J & = & 1 + 3k_2(a')^2 + 4k_3(a')^3 \cr\cr
& = & 1 + \frac{3b_3}{b_1}(a')^2 + \frac{2b_4}{b_1}(a')^3 \ . 
\label{j_sr3}
\eeqs
With this $S_{R,3}$ scheme transformation we find that for the
relevant range of $a \simeq 1$, $J$ can deviate excessively far from unity,
violating condition C$_1$. For example, for $a=1$ and $N=10$, we find that 
$J=339.8$, much larger than unity.  

One can also apply the various scheme transformations that we have devised in 
\cite{sch}-\cite{schi} to the beta function calculated in the 
$\overline{\rm MS}$ scheme and compare the resulting value(s) of the zero(s) of
the beta function with the value(s) obtained at the three-loop and four-loop
level in the $\overline{\rm MS}$ scheme.  Our general analyses in 
\cite{sch}-\cite{schi} (see also \cite{graceysch})
have shown that, for moderate values of the parameters 
determining these scheme transformations, the resultant values of the zero(s)
are similar to those obtained in the original $\overline{\rm MS}$ scheme.
In particular, the negative, unphysical value of $a_{rt,4\ell}$ will still be
present in the transformed scheme.  

Summarizing this section, we have shown that our conclusion, that the 
beta function of the finite-$N$ Gross-Neveu model, calculated up to four-loop
order, does not exhibit an IR zero, is robust with respect to scheme 
transformations.  


\section{Comparison with Results in the LN Limit and Behavior for $d > 2$}
\label{largeN_section}

In this section we discuss how the conventional perturbative beta function
reduces in the LN limit, and we also comment on some properties of the theory
for spacetime dimension $d > 2$.  From Eq. (\ref{lambda}), the quantity
that remains finite and nonzero in the LN limit is $\lambda = gN$, and hence
the corresponding beta function that is finite in this limit is
\beq
\beta_\lambda = \frac{d\lambda}{dt} = \lim_{LN} N \frac{dg}{dt} 
= \lim_{LN} N \beta \ . 
\label{betalambda}
\eeq
With the limit $N \to \infty$ having been taken, $\beta_\lambda$ has the 
series expansion, for $d \gsim 2$, with $\epsilon_d = d-2$, 
\beq
\beta_\lambda = \lambda \Big [ \epsilon_d + \sum_{\ell=1}^\infty \hat b_\ell
\xi^\ell \Big ] \ , 
\label{betalambdaseries}
\eeq
where
\beq
\xi = \lim_{LN} Na = \frac{\lambda}{2\pi} 
\label{xi}
\eeq
and
\beq
\hat b_\ell = \lim_{LN} \frac{b_\ell}{N^\ell} \ . 
\label{bellhat}
\eeq
Here we have used the fact that $b_\ell a^\ell = \hat b_\ell \xi^\ell$.  We 
find 
\beq
\hat b_1 = -2
\label{b1hat}
\eeq
and 
\beq
\hat b_\ell = 0 \quad {\rm for} \ \ell \ge 2 \ . 
\label{b234hat}
\eeq
The latter result follows from the fact that the structure of the bubble graphs
in the calculation of $b_\ell$ in, e.g., the $\overline{\rm MS}$ scheme, means
that, for $\ell \ge 2$, $b_\ell$ is a polynomial in $N$ of degree $\ell-1$.
Although the $b_\ell$ with $\ell \ge 3$ are scheme-dependent, this property is
maintained by scheme transformations that are finite in the LN limit
\cite{sch}. Hence, for $\ell \ge 2$, $\lim_{LN} b_\ell/N^\ell = 0$, which is
the result given in Eq. (\ref{b234hat}). Similarly, although $\hat b_\ell$ with
$\ell \ge 3$ are, in general, scheme-dependent, if they are zero in one scheme,
such as the $\overline{\rm MS}$ scheme, then they are also zero in any other
scheme reached by a scheme transformation function that is finite in the LN
limit \cite{sch}. It follows that in the LN limit, with 
$d=2+\epsilon \gsim 2$, 
\beq
\beta_\lambda = \lambda [ \epsilon - 2\xi ] = \lambda \Big [ \epsilon -
\frac{\lambda}{\pi} \Big ] \ . 
\label{betalambdaLN}
\eeq
Hence, 
\beq
d=2 \ \Longrightarrow \ \beta_\lambda = -\frac{\lambda^2}{\pi} \ , 
\label{betalambda_d2}
\eeq
with only the UV zero in this beta function at $\lambda=0$, and no IR zero. We 
can relate this to the beta function that was calculated in \cite{gn} in the LN
limit. From Eqs. (\ref{ggn}) and (\ref{betag}), we have
\beq
\beta = \frac{dg}{dt} = 2g_{GN} \, \frac{dg_{GN}}{dt} = 2g_{GN}\beta_{GN} \ .
\label{betabetagn}
\eeq
Explicitly, in the LN limit, from Eqs. (\ref{betalambda_d2}) and
(\ref{ggn}), 
\beq
\beta_\lambda = -\frac{\lambda^2}{\pi} = -\lim_{LN} \frac{g_{GN}^4 N^2}{\pi} \
.
\label{betarel}
\eeq
Combining Eqs. (\ref{betalambda}), (\ref{betabetagn}), and (\ref{betarel})
yields $\beta_{GN} = -g_{GN}^3 N/(2\pi) = -g_{GN}\lambda/(2\pi)$, in agreement
with Eq. (\ref{betagn}) above, or equivalently, Eq. (3.7) in Ref. \cite{gn}. 
This agreement was guaranteed, 
since the LN limit is a special limit of the result for finite $N$.
Accordingly, our finding that there is no robust evidence for 
an IR zero in the finite-$N$ beta function of the ($d=2$) Gross-Neveu model is,
{\it a fortiori}, in agreement with the fact that in the LN limit, the beta
function $\beta_\lambda$ in Eq. (\ref{betalambda_d2}) (equivalently, 
$\beta_{GN}$ in Eq. (\ref{betagn}) above), does not exhibit an IR zero.

If $d > 2$, then for small $\lambda$, the GN theory is IR-free, with an IR zero
of $\beta_\lambda$ at the origin, $\lambda=0$, and a UV zero of $\beta_\lambda$
at
\beq
\lambda_{UV}= \pi \epsilon \quad {\rm for} \ d \gsim 2, 
\quad {\rm LN \ limit} \ , 
\eeq
which is a UV fixed point of the renormalization group.  This is closely
analogous to the result found from an exact solution of the O($N$) nonlinear
$\sigma$ model (NL$\sigma$ M) in $d=2+\epsilon$ dimensions in the $N \to
\infty$ limit \cite{nlsm}.  In that theory, denoting the analogous finite
coupling in this limit as
\beq
x = \lim_{N \to \infty} N \lambda_{NL\sigma M} \ , 
\label{x}
\eeq
the exact solution yielded the beta function, for $d \gsim 2$,  
\beq
\beta_x = \frac{dx}{dt} = x\Big [ \epsilon - \frac{x}{2\pi} \Big ] \ .  
\label{betax}
\eeq
Thus, this nonlinear sigma model is, like the GN model in $d \gsim 2$, IR-free 
with a UV fixed point at 
\beq
x_{UV} = 2\pi \epsilon \ . 
\label{xuv}
\eeq
%


\section{Conclusions}
\label{conc}

The Gross-Neveu model in $d=2$ spacetime dimensions has long been of value as
an asymptotically free theory which is exactly solvable in the LN limit and, in
that limit, exhibits nonperturbative fermion mass generation and associated
dynamical chiral symmetry breaking.  In this paper we have considered the
finite-$N$ Gross-Neveu model. We have addressed and answered a fundamental
question about the UV to IR evolution of this model, as embodied in the beta
function, namely whether this beta function exhibits evidence for an IR
zero. For the purpose of our study, we have analyzed the beta function to the
highest-loop order to which it has been calculated, namely the four-loop
order. Our study used a combination of three methods, namely a direct analysis
of the three-loop and four-loop beta functions, a study of Pad\'e approximants,
and a study of the effect of scheme transformations.  We find that in the range
of coupling where the perturbative calculation of the four-loop beta function
is reliable, it does not exhibit robust evidence for an infrared zero.


\begin{acknowledgments}

This research was supported in part by the Danish National
Research Foundation grant DNRF90 to CP$^3$-Origins at SDU (T.A.R.) and
by the U.S. NSF Grant NSF-PHY-16-1620628 (R.S.)

\end{acknowledgments}


\newpage



\begin{thebibliography}{99}
\bibitem{indices}
%
Here the copy indices $j$ on the fermion field $\psi_j$ are implicit in the
notation, i.e., $\bar\psi \dslash \psi \equiv \sum_{j=1}^N \bar\psi_j \dslash
\psi_j$ and $(\bar\psi \psi )^2 \equiv (\sum_{j=1}^N \bar\psi_j \psi_j)^2$.
\bibitem{lsym}
%
Ref. \cite{dhn} showed that the GN model is actually invariant under a larger
global O($2N$) symmetry, of which $G_{cg}={\rm U}(N)$ is a subgroup, but we 
will not need this property for our present analysis.
\bibitem{otherbeta}
%
In our Eq. (\ref{betagn}) we are quoting the result from Eq. (3.7) of
Ref. \cite{gn}. An alternate calculation of $\beta_{GN}$ was given in
Eq. (4.16) of Ref. \cite{gn}, with the result $\beta_{GN} =
-(2\pi)^{-1}\lambda g_{GN}/[1 + (\lambda/(2\pi))]$. As was noted in \cite{gn},
their expressions for $\beta_{GN}$ in their Eqs. (3.7) and (4.16) agree to
leading order in $\lambda$.
\end{thebibliography}
\end{document}